# What Is Software Engineering? *

## Fedor Dzerzhinskiy, Leonid D. Raykov

***Abstract:*** A later translation (2015) of the article in Russian published in 1990. The article proposes an approach to defining a set of basic notions for subject area of software engineering discipline. The set of notions is intended to serve as a basis for detection and correction of some widespread conceptual mistakes in the efforts aimed at improving the quality and work productivity in creation and operation of software.

According to one of standardized definitions, "*software engineering*" is "*The systematic approach to the development, operation, maintenance and retirement of software*" [1]. The issues of research, development, and practical implementation of such "systematic approaches" constitute an important

---

Authors: Fedor Dzerzhinskiy, CSDP (fdzer@acm.org), Leonid Dmitrievich Raykov, PhD

area within the theory and practice of computing. This area, for which the English name "software engineering" became the main internationally accepted name (see, e.g., sources [1-11]), and which will be denoted here SE, exists and is being rapidly developed already for more than twenty years[1]. It is represented by a rich, constantly augmented body of published achievements, has "own" scientific and technical periodicals, such as *IEEE Transactions on Software Engineering* and *Software Engineering Journal*, regularly appears on pages of many other journals, magazines, proceedings, etc.

An analysis (based on available literature, samples of software, other sources) of concrete sum of achievements embodying the global state-of-the-art level of SE, allows to try to explicate the above-cited concise definition of SE by means of the following extended working definition:

*Software engineering* - concepts, methods, techniques, tools, systems, etc., designated for ensuring high quality and productivity of engineering activities on creation and operation (i.e., use) of software. (In other words, everything of what persons involved in such activities reasonably should know, should have a skill to do, and should have in disposition, in order to perform the work in a professional way.)

Let's agree to call the activities, mentioned in this definition, *software engineering activities* or *SE processes*.

We use here the notion of software in a wider sense, as covering not only "executable software" (computer programs), but also such objects as databases, knowledge-bases, and other, including "non-executable software"[2]. But for simplicity the reader may ignore this nuance and regard the software as only executable software.

Currently (1989) main part of works and publications in Russian on the issues of SE are related to the science and technology area known as "technology of computer programming" (or "technology of programming," TP) [12, 13]. These words in Russian are often used as a translation of English name of SE as a discipline (along with many other variants, including "designing of program systems" and "system programming", this is how "software engineering" was translated in the title of books [2, 5] published in Russian). But in fact they rather denote a specific "current" in the area of SE and, possibly, of some nearby areas. This current of SE discipline has it's own historical roots [14, 15], rather autonomous traditions, priorities distribution, a body of characteristic original results, own disputable points, etc. (see, e.g., works [16-21]). This "technology of programming" current in the discipline of SE we will denote here *SE-TP*.

An experience shows, that in the course of planning and performing various steps towards improving the practice of programming and use of computers, some typical conceptual mistakes

---

[1] F.D., 2015: The cited IEEE standard definition of SE as of 1983 [1] has been changed in 1990 [*55], and persists until now in the following wordings: "*software engineering. (1) the systematic application of scientific and technological knowledge, methods, and experience to the design, implementation, testing, and documentation of software*" [*56], "*(2) the application of a systematic, disciplined, quantifiable approach to the development, operation, and maintenance of software; that is, the application of engineering to software*" [*57]. (Cited from the SEVOCAB, http://www.computer.org/sevocab , visited August 2015).

The application of an approach to doing something, or a method of doing is not clearly distinguishable from simply doing this. And the name SE according to these new definitions is often regarded as simply a synonim of "software development."

Nevertheless the "systematic approaches," as a subject of development, study, and practice in their own right and value, didn't disappear with the change of the standard definition, they only became less convenient to mention. In order to resolve the umbiguity, let's agree to call the "old" meaning of SE as "SE1," and to the "new" meaning as "SE2." In this article SE means SE1, and instead of SE2 we mention and discuss software development.

[2] F.D., 2015: In this translation we use the terminology of "executable" and "non-executable" software as adopted in ISO/IEC19770-1:2006 [*58] and ISO/IEC19770-1:2012 [*59].

with quite costly and hardly-repairable consequences are being caused by the popularity of incomplete, inaccurate, inconsistent views on the essence of tasks, achievements, and characteristic features of current state of SE discipline. Such distorted views, in particular, can be a result of consideration of the SE issues through a prism of some approaches which have evolved historically in SE-TP.

A couple of trends in SE-TP appear to contribute most to propagation of such distorted, narrow views on SE discipline. These trends can be described non-rigorously as manifestations of the "machine-building" approach, the "administrative" approach, and the "tool" approach.

## Narrow Views on the SE-TP [3]

**1. The "machine-building" approach.** An essence of this approach consists in interpretation of idea of "technology of programming" by a literal enough analogy with the technological processes of production in the machine-building industry, while such processes are being regarded as strictly regulated sequences of "technological operations" which have to ensure the creation of final product having the specified properties, with the minimum dependence on individual abilities (and, especially, on creative abilities) of separate workers, who participate in execution of "technological operations."

One should notice, that not only in programming, but in machine-building too, not any sort of work can be effectively executed with the use of this approach. A necessary condition of applicability of this approach is the possibility to split up the work - in an acceptable way, in advance, prior to actually doing the main part of the work - into a set of strictly and stably defined, repeatedly executed, narrowly specialized operations.

Within the entire cycle of development and production of machines, the mass manufacturing of products is a typical example of works, which as a rule satisfy this requirement. Examples of works that usually do not satisfy this requirement: the products development; the inventor activities; the research for creation of new products.

In the field of software engineering activities also there are sorts of work which satisfy this requirement. Such work includes mainly the following: trustworthy copying of the data, computer media, and documentation; the simplest, routine forms of setting up and generating the programs; some narrowly specialised software development tasks, being performed without a significant degree of novelty, within a rigid, repeatedly tested schema, the applicability of which is being achieved due to limitation of variety of application problems being solved. Such sorts of work sometimes are called *production* or *manufacturing* of software.

Not belittling a practical significance of these sorts of work, we'd like to emphasize, that they are not at all the same as the most characteristic works in the SE field, the processes of new software development (like how the typography processes, an industrial production of copies of the book, are normally not a part of the work of authors who write books, create the books as an information). Any of the above-mentioned software "manufacturing" processes require prior development, in some way, of the software which will be "manufactured" (i.e., will be copied, set up, will constitute a basis of the solution scheme for a certain class of application problems, etc.).

As regards the processes of new software development, these processes, within an analogy with the machine-building, naturally correspond also to the development, not to the production of machines. And thus the very discussion of applicability of "machine-building" approach to such processes is hard to explain otherwise than as a terminology imprecision, when software development is called production in order to make it sound "more impressive."

---

[3] F.D., 2015: This section of the article was intended for a Russian reader of the late 1980s. Some of the material in it can seem not quite up-to-date in 2015. The more busy readers can be recommended to skip this section, or take a look at it selectively.

Also, there is a direct contradiction between some widely recognized achievements in the field of SE and the pre-defined, fine-grained breakup of work processes into "technological" operations, which is characteristic of the "machine-building" approach. Such a breakup, in particular, violates the principle, known since the appearance of the structured programming discipline [22], that in the process of top-down stepwise refinement of the program logic, one should as much as possible take at each step only a single essential technical decision, and at the same time try to postpone each decision until the decision becomes really necessary (the later the decision is taken, the more chances that it will be satisfactory; a premature decision limits the possibility of choosing optimal solutions at subsequent steps of the development).

In practice, attempts to apply the "machine-building" approach in the development of new software often lead to bureaucratism of the work processes, make them overly conservative, highly inertial. Also, even when such attempts, undertaken by a software developers team, seem to be not evidently ineffective, they may have the postponed negative consequences, lead to growing methodological retardedness of such team by hampering adequate perception of the global state-of-the-art level of SE.

**2. *The "administrative" approach.*** This approach to achieving the goals of SE emphasizes the role of various administrative decisions and directives, related to the enforced adoption of standards, rules, internal regulations, of the choices among alternative software platforms, software solutions, etc.

When preparation and implementation of such decisions is being performed in a professional manner, on the basis of deep and competent analysis of all relevant technical aspects, they can be very effective. Unfortunately, at present, often enough one can encounter the attempts to implement such decisions too hastily and on an overly wide scale by the administrative enforcement methods.

Neglecting the necessity of meticulous analysis, performed on the basis of professional software developers experience, of possible consequences of such decisions, which is typical of some managers and officers involved with such decisions, can lead to unrestrained growth of bureaucracy of processes, to over-staffing, to formation of exaggerated, unproductive "management superstructure" above the SE processes.

It's worth noticing, that in such "extreme" manifestations of administrative approach the most serious negative consequences are often caused not by a too strict directive regulation of the work, but quite contrarily - by the fact that inability to practically ensure the compliance with the imposed formal rules leads to the loss of respect for subsequent directives, highly complicates the adoption of the strict rules where these are really necessary and otherwise would be achievable. This resembles "breaking the screw thread" as a result of excessively "tightening the screw."

**3. *The "tool" approach.*** A quite popular interpretation of SE-TP is it's one-sided interpretation as some software tools. In "extreme" manifestations this approach leads to the attempts of invariably taking into account only those solutions of any problems, related to software development productivity and quality, which consist in the development or adoption of some software tools - from the simplest to the integrated work environments, etc. Various software products and systems of this kind (the software designed for creating other software with its help) are commonly called *software tools* [23, 24].

Of course, what deserves objections regarding the "tool" approach, is not an emphasize of the importance of software tools, but an attempt to reduce the entire concept of SE discipline to mainly or solely this single aspect. Such interpretation of SE carries an inherent illusion that the perfection of professional mastery of the specialists, the most important, creative part of their work, can be replaced by means of implementing additional automated functions of control, communication and help. This illusion can have severe negative consequences.

First, it should be noted that the popularity of such an illusion causes mass misunderstanding and limited use by the specialists of that key achievements of the SE and SE-TP, which are of *conceptual* (*methodological*) nature. Examples of such achievements are the current best practices of software user documentation development [25-27], software requirements specification [28, 29], software quality assurance [30, 31], principles and techniques of practical structured programming [22, 32-34], etc.

Disregarding such achievements can lead to a lower quality of the very software tools being created or adopted. In some cases, these tools, being useful in many respects, or at least in some respects, at a closer look reveal that they allow too narrow class of operational environments, are not enough reliable, not easy to use, unwieldy, require disproportionate resources, have incomplete or incomprehensible documentation. Additional negative side effect, related to the use of such imperfect tools, is that "ordinary" developers willingly or unwillingly regard their tools as an example of "normal" level of software quality, and get used to regarding the results of their own work uncritically too.

One more negative consequence of the "tool" approach in SE-TP can be characterised as a disorientation of some of the application software developers regarding the practical steps which at a current moment are required of them (not of software tools developers) in order to improve their productivity and quality of their work results. In the absence of perfect "productivity" tools, postulation of the key role of such tools leads those, who should have used the tools, to either the attempts to create their own tools, possibly in a semi-dilettante manner, or (more often) to justification of their imperfect, outdated methods of work by references to the absence of tools, or, finally, to underproductive attempts of adopting imperfect or unsuitable tools.

And if the necessary tools are in principle available to their potential users, then the "tool" approach still suggests a disputable idea of adopting as much powerful software tools as possible, not taking into account the related costs (of tools licensing, adoption, etc.), the possibility of achieving acceptable results with lower expenses, not necessarily related to adoption of these new powerfull software tools. In relation to the above, it is necessary to note, that globally, despite the significant achievements in the development of CASE (Computer-Aided Software Engineering) systems, the applicability of such systems is not at all being regarded as unconditional - because of high costs of necessary software tools [35].

## Software Engineering Resources

Let us return to the wider interpretation of the SE subject, as introduced above, and continue to discuss on this basis the factors, which appear to restrain the evolution and application of achievements in the SE area. One such factor is the absence, at present, of recognized criteria to distinguish the finalized, complete forms of achievements in the area of SE from their preliminary or intermediate forms, when considering the creation, promotion, adoption, systematic practical use, and the evaluation based on the results of use of that or other such achievement.

Let's agree to call the desirable completed form of SE achievements "SE discipline resources" or simply "SE resources." Below we'll try to consider some characteristic features of such a "resource," and also mention negative consequences of imitating it by two other forms of achievements - concrete practical experience and research results.

We will define an *SE discipline resource*[4] (shortly, *SE resource*) as a collection of some results pertaining to the area of SE, which are of "conceptual" nature (methodological, organizational, etc.),

---

[4] F.D., 2015: In Russian - "*sredstvo*".

consist of some software, or have some other nature, and possess the following distinctive properties:

Firstly, repeated, immediate applicability in a certain class of working conditions;

Secondly, reliable enough reproducibility, when applied properly, of certain practical effect (actually or presumably useful);

Thirdly, identifiability - possibility to distinguish objectively enough and accurately this concrete resource from any other while the resource is considered as an object of application, adoption, transfer, etc.

Because of significant natural variability of any concrete environments of application of an SE resource (changes in the composition of equipment, personnel, of the tasks being solved, etc.), a repeated applicability of the resource is not achievable without some degree of its *transferability* (possibility to pass the resource from one party or environment to other). Thus the transferability is an important attribute of a high-quality SE resource. (The notion of SE resource includes also the low-quality resources.)

Some examples of SE resources:

- "organizational and methodological" resources - the IEEE Software Engineering Standards series [1, 36 - 40];

- "pure-software" and "software-conceptual" resources - various software tools and environments supporting software development processes [24];

- "pure-conceptual" resources ("*method-resources*") - various versions of the Software Walkthrough techniques [33, 41], similar-purpose technique of the "*Structured Inspection*" ("*Structured Expertise*") [42].

The proposal to consider SE resources as a desirable completed form of the SE discipline's achievements is not aimed at belittling the value of two other typical and necessary forms of such achievements - the concrete practical experience (systems, created "internally" in some concrete organization for quality execution of activities related to programming) and the research results. At the same time it is important to emphasize wrongness of popular enough attempts to judge about the SE achievements, as if they were a concrete experience or a research result, in situations, where essentially these achievements have to perform the role of SE resources (for example, when the achievement is being proposed for a wide practical adoption).

Interrelation between research results, concrete experience, and an SE resource is similar to that between an initial concept and algorithm sketch of some computer program, "simply" a program, used by its author in the environment for which it was developed, and a universalized, fully tested, documented, etc. "software product" (according to Brooks [2]).

Research results, as a rule, are not immediately applicable in practice, and this is not expected of them. (Their application requires serious additional efforts, solution of various technical, organizational and other issues). For example, concepts of Structured Prorgamming (a research result) are being detailed and interpreted, in order to make them immediately applicable in "production" software development, in many different ways, among which there are both very productive and unpractical ones.

A concrete practical experience is usually not transferable, because it depends too much on the unique specifics of the organization which has accumulated it. More than this, it is often impossible to know for sure, what really generates the useful effect achieved in this experience - the formal and technical attributes of the processes, most visible superficially, or the personal involvement of certain especially talented individuals?

Taking into account the above, it makes sense, while paying the due to SE research and to studying and propagation of advanced experience, to concentrate attention on creation and dissemination of SE resources - high-quality, repeatedly tested, immediately applicable.

# Organizational Sociotechnical Systems and Sociotechnical Environment of Software Engineering

Application of any SE resources can be viewed as an organizational effort, which is aimed at improving some SE processes, and consists in affecting purposefully, in some manner, the sociotechnical systems[5] in which these processes are being executed. It is useful to consider the form which this affecting can generally take in practice.

An *Organizational Sociotechnical system* (*OST-system*) - will mean here an individual or a group of individuals, united by joint activities, and being regarded as an organized whole ("*organization*") together with some concrete resources, conditions, and rules of these activities. As applied to the SE activities, the resources, conditions, and rules usually comprise computers and other equipment, information and other resources, interconnections with other OST-systems, etc.

An *Organizational Sociotechnical environment* (*OST- environment*) - will mean everything, in the concrete OST-system performing some SE processes, that influences productivity and results' quality of these processes. Typically, it is necessary to take into account the existence of at least the following important constituents of the OST-environment:

- the equipment (first of all, certain configurations of computing devices and systems);

- the software - specific versions, installed, familiar to personnel, and ready for use;

- regulatory documents (legal, normative, internal, and other) in force with regard to this OST-system;

- other documentation, literature, information sources, used by the OST-system's personnel in their work;

- individual and collective knowledge, skills, experience, habits, traditions, possessed by this personnel (hereunder, for shortness - *personnel's knowledge*).

This list is not meant to be complete. (For example, one could also append to this list everything of what is influencing the general work conditions of the personnel - convenient work regime, availability of computing and communication resources, psychological climate in the team, etc.). However, what is listed above is already sufficient for an important practical conclusion: only a minor part of constituents of the OST-environment can be directly changed as a result of control actions such as inclusion (acquisition) of some "productivity" software tools, issuance of internal governing documents, etc. And the "net effect" of such control actions, which is determined indirectly, through the changes of other constituents of the OST-environment (especially, through the changes of the personnel's knowledge), can be quite far from what was the intention.

Because of the same reason (the significant inter-dependency of the OST-environment constituents, while the personnel's knowledge has the key role), a goal of "creating" an OST-environment with specified properties, is remaining this in practice only until the composition of principal participants of SE activities is determined (the "kernel" of the OST-system personnel). After this, the individual and collective knowledge of this personnel come into action, the process of some kind of "self-organization" of the OST-environment begins, and the tasks of ensuring any specified properties of it can be solved *only by means of changing* the OST-environment.

---

[5] FD, 2015: In the original Russian version of the article the "organizational and technical" ("*orga-technical*") systems are being mentioned. But the intended meaning exactly corresponds to current notion of sociotechnical systems, the fundamental role of which with regard to IT and computing is being pointed out in works such as [*60].

All the above allows to suggest the following principle conclusion regarding the nature of SE resources. Any SE resources, when being applied in practice, always constitute the means of changing (not creation "from scratch") of the SE OST-environments, and changing indirectly, through mediation by the OST-system personnel's knowledge. Disregarding of this, unfortunately rather popular, significantly complicates adequate prediction and management of the SE resources use's net effect.

## Efficiency of SE Resources

We regard the notion of SE resource as referring not to some rigorously defined class of high-quality SE achievements, but to a role, for which quite diverse achievements can be candidates, reasonably or not, and which should aid evaluation in a systematic way, in comparable categories of the quality and efficiency of performance in this role of various achievements.

In other words, in any cases, when some achievement is being considered, which due to a way of its actual or expected use may potentially perform the role of SE resource, it makes sense to ask not a terminological question of whether or not this is a "resource," but *how good or bad* an SE resource is this achievement, what efficiency it will possess when applied in this role?

An *efficiency* of the SE resource is understood here as an integral characteristics of the socially and economically significant effect of its application, which takes into account the total repertoire of both positive and *negative* aspects of this effect. This notion is related to one of key ideas in the discussed approach to SE issues, namely, that the consideration of application, development, and, especially, of administratively enforcing the use of any SE resources can and should take into account not only positive outcomes, supposedly ensured in the case of adoption of these resources, but also the cost of achieving these outcomes. And, aside from direct expenses, one should consider:

- the availability of less expensive ways of achieving equivalent or better results;

- how much well-grounded, in the present circumstances, is an intention to achieve the outcomes, which are expected of the SE resource (and not some less ambitious outcomes, but with the significantly lower expenses);

- the risk that the SE resource, being proposed for adoption, even if it will provide certain positive effect, will become an obstacle to implementation of other, more effective and necessary changes, etc.

Application of typical SE resources in many cases requires quite sensible expenses, which can be regarded as justified costs by far not always. More than this, not insignificantly rare are situations, when the expenses many times exceed the obtained useful effect (see, for example, [43, 44]).

Quantitative estimation of the SE resources efficiency is a separate subject, which is not discussed in this paper. There are many publications on this subject, including the works [44 - 47]. Here we'd like to touch general qualitative consequences of the proposed approach to defining the efficiency of SE resources. In order to illustrate the character of these consequences, we will discuss as an example one of them, related to the notion of SE resources modularity.

## Regarding Modularity of SE resources

The above idea of SE resources as a means of changing the OST-environment, and not of creating it anew, helps to notice, that a common source of losses of the SE resources efficiency are various inconsistencies between the content of changes being introduced, on the one hand, and the already accumulated SE achievements, or other peculiar aspects of the OST-environment, on the other hand.

If an SE resource introduces into the OST-environment, along with rational changes of one aspect of it, the non-rational changes of other aspect (for example, if it conflicts with the corresponding elements of some other SE resource, adopted previously and more perfect in this aspect), then this results in either "forcing out" previous useful experience, or the rejection of the new SE resource.

In its turn, wide dissemination of SE resources, and, in this way, accumulation of maximally diverse experience of their practical use, is a necessary pre-requisite to a systematic improvement of the best elements of such resources, to "filtering out" their less successful elements, and thus to continuously selecting, in a way of competition, the best elements of various SE resources, while providing for the possibility of their productive joint use.

Finally, in most real-life situations a radical, one-time change of an SE OST-environment is equivalent to total dis-organization of it (the loss of all of previous experience and groundwork for future tasks). This leads to choosing partial and gradual (phased) ways of performing such changes.

All the above makes preferable that the collections of SE resources shall have *modular organization*, which is understood here as an organization of contents of these resources, such that this content is represented as some hierarchy of separate fragments (modules), composed so that, as far as possible:

- all fragments should be maximally suitable for their selective learning and application (inapplicability or inefficiency of some fragments in certain work conditions should not hinder application of other fragments that can be productively applied in these conditions);

- the total collection of fragments should allow continuous evolution of it by way of independent enough improvement of constituent parts, and maximum freedom of adding new, potentially useful fragments (including independently developed and/or alternatives of the already included fragments having the same purpose).

At present, in the literature there are mentionings of enough programs or projects in the area of SE resources development which employ modular approaches similar to the above. Some examples are: U.S. DoD STARS program (Software Technology for Adaptable, Reliable Software) [48]; IBM Software Engineering Support Facility [49]; the conception of development of SE-TP in the The comprehensive programme of scientific and technical progress of the CIS member-states [13].

It should be noted, that the reports about such projects sometimes are interpreted by specialists in a one-sided way, as dealing with the modularity only with regard to those SE resources which constitute some software tools. Such interpretation leads to emphasizing the technical issues related to creation of standardized platforms supporting the compatibility of separate software tools at a level of internal data representation, and the like. This differs from the idea of SE resources modularity, as described above. In relation to this, the following can be pointed out.

Firstly, the above argumentation in favor of SE resources modularity is applicable to any sorts of them, and specifically to the conceptual resources ("*method-resources*"), such as techniques, recommendations, technical agreements, including standardized ones, etc., not only the software tools.

Secondly, like the use of a programming language, that supports modularity, can only facilitate, but not ensure achieving the modularity of programs created, as much the software development platforms designed to support the software tools modularity, can support it effectively only in the case, when the formal conventions regarding modules interconnection, data structures, etc. (the public tool interface [50]) are being used in combination with a certain sum of *conceptual* *components*, such as concepts frameworks, principles, rules, techniques, technical policy, aimed at ensuring the modularity of SE resources at the substantial, non-formal level.

<p style="text-align:center">*   *   *</p>

Returning to the question, which forms the title of this article (what is SE?), we'd like to notice this. As a proposed variant of answer to this question the general considerations, discussed here, are intended for use in combination with available information about the concrete characteristic elements and achievements which determine the current state of SE as an area of study and practice. The reader can get such information in the literature[6], including the sources cited above. Classification and review of achievements in the SE area according to the proposed vision of this area is a separate question, that is not discussed here.

Discussion of some details and additional aspects of the proposed view on the subject of SE discipline can be found in publications [51 - 54].

---

[6] F.D., 2015: Presently there are a lot of Bodies of Knowledge (BoK) or Curriculum Guidelines which systemize these issues. Two such sources are SWEBOK Guide 2004 [*61] and GSwE2009 [*62]. Both classify the following knowledge areas (KAs) as pertaining to SE: Software Requirements, Design, Construction, Testing, Maintenance, Configuration Management, SE Management, SE Processes, Software Quality. SWEBOK classified the Tools and Methods as a separate KA. In GSwE2009 three important additional KAs are included and characterized: Ethics and Professional Conduct (including Legal Issues), Systems Engineering, and Engineering Economics ([*62], Appendix C). Some of this and some more was added in 2014 version 3.0 of SWEBOK Guide [*63]. In [*61 - *63] one can find a lot of useful references to literature. (In this respect, new version of SWEBOK Guide [*63] does not totally replace an old version [*61], but rather complements it.)

Also today in practice an indispensable aspect of SE activities are the issues of Security (including Privacy, Data Protection, and, when applicable, Safety as a special crucial concern). This is so much important, that usually is being evolved and considered as a separate area in its own right. There are lots of sources, of which we'll mention only a few examples - [*64 - *67]. In PCI DSS [*66], the Requirement 6 "Develop and maintain secure systems and applications" can serve as one illustration of "intertwining" relationship between Security and SE. There are abundance of other examples - in Communications, Medicine, Avionics etc.

———————